\documentclass{elsart}
\usepackage{hyperref}
\usepackage{amssymb}
\usepackage[dvips]{graphicx}% Include figure files
\usepackage{dcolumn}% Align table columns on decimal point
\usepackage{bm}% bold math
\usepackage{amsmath}
\renewcommand{\vec}[1]{{\mathbf{#1}}}
\newcommand{\beq}{\begin{eqnarray}}
\newcommand{\eeq}{\end{eqnarray}}
\begin{document}

\begin{frontmatter}


\title{Mottness}

\author{Philip Phillips}

\address{Department of Physics, University of Illinois, Urbana-Champaign, Illinois, 61801-3080}

\begin{abstract}
We review several of the normal state properties of the cuprates in an attempt to establish an organizing principle from which pseudogap phenomena, broad spectral features, $T-$linear resistivity, and spectral weight transfer emerge. 
We first show that standard field theories with a single critical length scale cannot capture the $T-$linear resistivity as long as the charge carriers are critical.  What seems to be missing is an additional length scale, which may or may not be critical. Second, we prove a generalised version of Luttinger's theorem for a Mott insulator.
Namely, regardless of the spatial dimension, the Fermi surface of the non-interacting system is converted into a surface of zeros of the single-particle Green function when the Mott insulator posesses particle-hole symmetry. Only in the presence of particle-hole symmetry does the volume of the surface of zeros equal the particle density.  The surface of zeros persists at finite doping and is shown to provide a framework from which pseudogaps, broad spectral features, spectral weight transfer on the Mott gap scale can be understood.  
\end{abstract}
\end{frontmatter}

\section{Introduction}

The normal state of the high-temperature copper-oxide superconductors (cuprates for short) is anything but normal.  First, all parent cuprates are Mott insulators\cite{mott}.  Such insulators do not insulate because the band is full, but rather because strong local electron correlations dynamically generate a charge gap by splintering a half-filled band into lower and upper Hubbard bands. Second, once doped, Mott insulators still insulate.  In fact, numerous experiments\cite{boeb1,boeb2,boeb3,boeb4} in which a large magnetic field is applied to kill superconductivity reveal that the resistivity diverges as the temperature tends to zero below a critical doping level, $x_c$.  Above $x_c$, traditional Fermi liquid behaviour is observed in the overdoped regime.   Third, electron quasiparticles are nowhere to be found in the underdoped regime where the ubiquitous 
pseudogap\cite{alloul,norman,timusk} phenomenon obtains.  Fourth, near optimal doping, the resistivity scales as a 
linear\cite{batlogg} function of temperature as opposed to the quadratic behaviour indicative of a Fermi liquid.  Such oddities did not, of course, beset a theoretical understanding of
superconductivity in metals such as aluminum because the normal state is accurately rooted in Fermi liquid theory.  In the cuprates, the central problem of the normal state is understanding how the disparate phenomena just described arise from strong electron correlations intrinsic to
 the Mott state.  
Simply stated, the question is: is there a general organizing principle that captures Mottness?  Mottness refers to those features of doped Mott insulators that are not deducible from ordering.  While it might be that ordering is deducible from Mottness, Mottness is distinct nomologically from ordering.  Indeed, it is Mottness that makes the cuprate problem largely intractable.  It is for this reason that numerous theories of the cuprates get Mottness clearly wrong, ignore it, or deny its existence.   In this paper, we prove two theorems which bare directly on the nature of the low-energy theory for a doped Mott insulator. Both theorems point to physics beyond that which is captured by standard field theories based on one-parameter correlation length scaling.  In the first theorem, we show explicitly that within single parameter scaling, quantum criticality cannot account for the linear $T$ resistivity.  Second, we show that for Mott insulators possessing particle-hole symmetry, the single-particle Green function vanishes at the Fermi surface of the non-interacting system.  While the surface of zeros is shown to persist in the absence of particle-hole symmetry,
only in the presence of such a symmetry does the volume of the surface of zeros equal the particle density.  The surface of zeros provides a framework from which pseudogaps, broad spectral features, and spectral weight transfer on the Mott gap scale can be understood.  

\section{T-Linear Resistivity}

One of the leading explanations for the $T-$linear resistivity is quantum criticality\cite{qc1,qc2}. We do not analyse here specific proposals such as the highly successful marginal-Fermi liquid theory\cite{varma}, which is in principle consistent with a quantum critical point, but rather dissect the assumptions of quantum criticality which lead to $T-$ linear resistivity.  At the quantum critical
coupling, or in the quantum critical regime, $k_B T$ is the only energy scale governing
collisions between quasiparticle excitations of the order parameter.  Consequently, the transport relaxation rate,
\begin{eqnarray}
\label{lint} 
\frac{1}{\tau_{\rm tr}}\propto\frac{k_BT}{\hbar}, 
\end{eqnarray} 
scales as a linear function of temperature, thereby implying a $T-$linear resistivity if (naively) the scattering rate is
what solely dictates the transport coefficients. That temperature is the
only scale in the quantum critical regime regardless of the nature of the
quasiparticle interactions is a consequence of universality. Eq. (\ref{lint})
holds as long as the inequalities $T>|\Delta|$ and
$t<1/|\Delta|$ are maintained, $\Delta$ the energy scale measuring the
distance from the critical point and $t$ the observation time. The energy scale $\Delta\propto \delta^{z\nu}$ varies as a
function of the tuning parameter $\delta=g-g_c$, where $\nu$ is the
correlation length exponent and $z$ is the dynamical exponent. At the
critical coupling $\delta=0$ or $g=g_c$, $\Delta$ vanishes. Ultimately, the
observation time constraint, $t<1/|\Delta|$ implies that only at the quantum
critical point does the $T-$linear scattering rate obtain for all times. That
the quantum critical regime is funnel-shaped follows from the inequality
$T>|\Delta|$.
The funnel-shaped critical region should be bounded by a temperature $T_{\rm
  upper}$ above which the system is controlled by high-energy processes. Whether or not 
quantum criticality is operative up to temperatures of order $T=1000K$ in the
cuprates is questionable. 
\begin{figure}[t]
\centering
\includegraphics[width=7cm]{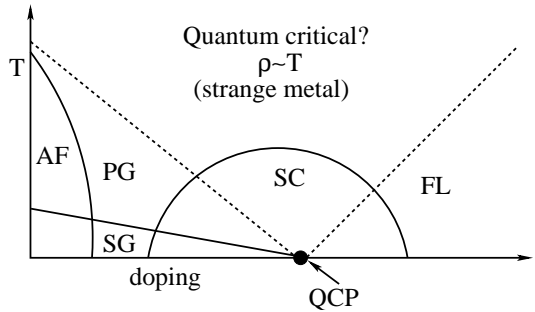}
\caption{Heuristic phase diagram of the cuprate
  superconductors as a function of temperature and hole doping level. 
 AF represents antiferromagnet, SG, the
  spin glass, SC the superconductor, PG the pseudogap regime
in which the single-particle specrum develops a dip and FL the Fermi liquid. The spin-glass phase ceases at a
  critical doping level (quantum critical point, QCP) inside the dome.  The
  dashed lines represent crossovers not critical behaviour.  The strange-metal behaviour,
  $T-$linear resistivity, in the funnel-shaped regime has been attributed to
  quantum critical behaviour.  A scaling analysis of the conductivity at the
  quantum critical point rules out this scenario, however.}
\label{fig1}
\end{figure}
 
We have recently formulated\cite{scaling} a general scaling argument for the conductivity 
to understand if $T-$linear resistivity is consistent with the general claims of quantum criticality.  The argument is based on three general assumptions:
1) the critical degrees of freedom carry the current, 2) only a single length scale is relevant near the critical point and 3) charge is conserved.  Granted true, these assumptions rule out quantum criticality as the cause of the $T-$linear resistivity.  

Before we present the proof, consider as a precursor to the Drude formula for the conductivity,
\begin{eqnarray}
\label{drude}
\sigma_{\rm Drude}=\frac{1}{4\pi}\frac{\omega^2_{\rm pl}\tau_{\rm
    tr}}{1+\omega^2\tau^2_{\rm tr}}, 
\end{eqnarray} 
with $\omega_{\rm pl}$ the plasma frequency and $\tau_{\rm tr}$ the transport relaxation time.  If we substitute Eq. (\ref{lint}) into Eq. (\ref{drude}), $T-$linear resistivity follows in the zero-frequency limit. However, explicitly
appearing in the formula for the resultant conductivity is the $\omega_{\rm pl}$. While the plasma frequency is commonplace in typical transport theories of metals, we show below that such a degree of freedom does not exist in quantum critical theories. Consequently, the predictions of Drude theory are not compatible with those of quantum criticality.  That inconsistencies lie with the Drude formula and quantum criticality can be seen immediately from the work of van der Marel\cite{vandermarel} and co-workers who used Eq. (\ref{drude}) to collapse their optical conductivity data as a function of $\omega/T$. True quantum critical behaviour would be described by an optical conductivity of the form, $T^{-\mu} f(\omega/T)$.  Contrastly, they find\cite{marel} that $\mu=1$ for $\omega/T<1.5$
and $\mu\approx 0.5$ for $\omega/T>3$ if $f(\omega/T)$ is described by the Drude formula.

Consider a general action $S$ whose microscopic
details are unimportant as long as the current is critical.  An externally applied electromagnetic
vector potential $A^\mu$, $\mu=0,1,\dots,d$, couples to the electrical
current, $j_\mu$, such that
\begin{equation} 
S\to S+\int d\tau \;d^dx\; A^\mu\;j_\mu.
\end{equation} 
Under the one-parameter scaling hypothesis for quantum systems, the spatial
correlations in a volume smaller than the correlation volume, $\xi^d$, and
temporal correlations on a time scale shorter than $\xi_t\propto\xi^z$ are
small, and space-time regions of size $\xi^d\xi_t$ behave as independent
blocks.   Using this hypothesis, we write the scaling form for the
singular part of the logarithm of the partition function by counting the
number of correlated volumes in the whole system:
%\begin{equation} 
%\ln Z=\frac{L^d\beta}{\xi^d\xi_t}F(\delta\xi^{d_\delta}, A^i_\omega
%f(\omega\xi_t)\xi^{d_A})
%\;,
%\label{eq:lnZ}
%\end{equation} 
\begin{equation} 
\ln Z=\frac{L^d\beta}{\xi^d\xi_t}\;F(\delta\xi^{d_\delta}, 
\{A^i_{\lambda}\; \xi^{d_A}\})
\;,
\label{eq:lnZ}
\end{equation} 
In this expression, $L$ is the system size, $\beta=1/k_BT$ the inverse
temperature, $\delta$ the distance from the critical point, and $d_\delta$
and $d_A$ the scaling dimensions of the critical coupling and vector
potential, respectively. The variables $A^i_{\lambda}=A^i(\omega=\lambda
\xi_t^{-1})$ correspond to the (uniform, $k=0$) electromagnetic vector
potential at the scaled frequency $\lambda=\omega\xi_t$, and $i=1,\dots,d$
labels the spatial components. Two derivatives of the logarithm of the
partition function with respect to the electromagnetic gauge $A^i(\omega)$,
\begin{eqnarray}
\sigma_{ij}(\omega,T)&=&
\frac{1}{L^d\beta}\;\frac{1}{\omega}\frac{\delta^2\ln Z}
{\delta A^i({-\omega}) \delta A^j({\omega})} 
\nonumber\\
&=& \xi^{-d}\;\frac{\xi_t^{-1}}{\omega}
\;
\;\xi^{2d_A}\;\frac{\delta^2}{\delta A^i_{-\bar\lambda}\delta A^j_{\bar\lambda}}
F(\delta=0)\Big|_{\stackrel{\bar\lambda=\omega\xi_t}{\{A^i_\lambda=0\}}}
\nonumber\\
&=&\frac{Q^2}{\hbar}\;\xi^{2d_A-d}\;\Sigma_{ij}(\omega\xi_t), 
\end{eqnarray} 
determine the conductivity for carriers with charge $Q$. We have explicitly
set $\delta=0$ as our focus is the quantum critical regime. At finite
temperature, the time correlation length is cutoff by the temperature as
$\xi_t\propto 1/T$, and $\xi_t\propto \xi^z$. The engineering dimension of
the electromagnetic gauge is unity $(d_A=1)$. From charge conservation, the current operators cannot acquire an anomalous dimension; hence, 
$d_A=1$ is exact~\cite{Wen1992}. We then arrive at the general scaling form
\begin{equation}
\label{genscal}
\sigma(\omega,T)=\frac{Q^2}{\hbar}\;T^{(d-2)/z}
\;\Sigma\left(\frac{\hbar\omega}{k_B
    T}\right) 
\end{equation} 
for the conductivity where $\Sigma$ is an explicit function only of the
ratio, $\omega/T$. (The $ij$ tensor indices have been dropped for simplicity.)
This scaling form generalizes to finite $T$ and $\omega$ the $T=0$ frequency
dependent critical conductivity originally obtained by Wen~\cite{Wen1992}.
The generic scaling form, Eq.~(\ref{genscal}), is also in agreement with that
proposed by Damle and Sachdev\cite{ds} in their extensive study of
collision-dominated transport near a quantum critical point (see also the
scaling analysis in Ref.~\cite{book1}). What the current derivation lays
plain is that regardless of the underlying statistics or microscopic details
of the Hamiltonian, be it bosonic (as in the work of Damle and
Sachdev\cite{ds}) or otherwise, be it disordered or not, the general scaling
form of the conductivity is unchanged.  The Anderson metal-insulator
transition in $d=2+\epsilon$, which can be thought of as a quantum phase
transition where the dimensionless disorder strength is the control
parameter~\cite{Wegner,Gang4}, obeys the scaling function derived here for the conductivity.

At zero frequency, the dc limit, we obtain
\begin{eqnarray}\label{dclimit}
\sigma(\omega=0)=\frac{Q^2}{\hbar}\;\Sigma(0)\;\left(\frac{k_BT}{\hbar
    c}\right)^{(d-2)/z} .
\end{eqnarray} 
In general $\Sigma(0)\ne 0$. Else, the conductivity is determined entirely by the non-singular and hence non-critical part of the free energy.
The cuprates are anisotropic 3-dimensional systems.  Hence, the relevant dimension for the critical modes is $d=3$ not $d=2$. In the latter case, the temperature prefactor is constant. For $d=3$, we find that $T-$ linear resistivity obtains
only if $z=-1$.  Such a negative value of $z$ is unphysical in standard commutative\cite{ncomm} field theories as it implies
that energy scales diverge for long wavelength fluctuations at the critical point. 

While it is certainly reasonable to assume that $\Sigma(0)$
  is finite at zero temperature, it is certainly a possibility that
  $\Sigma(0)$ might in fact diverge.  A possible scenario of how this state
  of affairs might obtain is that a dangerously irrelevant operator could
  govern the conductivity. In this case, $\Sigma(0)\sim 1/T^p$, and $T-$linear
  behaviour obtains if $p=(d-2)/z + 1$.  This can only occur above the upper
  critical dimension.  In this regime, all criticality is 
  mean-field-like.  Hence, a possibility that cannot be eliminated at this
  time is that all criticality in the cuprates is inherently mean-field and
  dangerously irrelevant operators control the conductivity.  For strongly
  correlated systems, however, it is unlikely that criticality is inherently
  mean-field.

Consequently, three options are available at this point: 1) quantum criticality has nothing to do with the problem, 2) the current is carried by non-critical degrees of freedom or 3) new quantum critical scenarios in which additional length scales describe the physics.  In
a scenario involving non-critical degrees of freedom, fermionic charge
carriers in the normal state of the cuprates could couple to a critical
bosonic mode. Such an account is similar to that in magnetic
systems\cite{qc1} in which fermions scatter off massless bosonic density or
spin fluctuations and lead to an array of algebraic forms for the
resistivity\cite{moriya,rosch} ranging from $T^{4/3}$ to $T^{3/2}$ in
antiferromagnetic and ferromagnetic systems, respectively.  While disorder
can alter the exponent\cite{rosch}, $T-$linear resistivity results only in a
restricted parameter space.  The robustness of $T-$linear resistivity in the cuprates makes this type of scenario unlikely.  What about new scenarios?  Let us entertain the possibility that an additional length 
$\tilde\xi$ is relevant and diverges as $\tilde\xi\propto \xi^a$, with
$a>1$.  In the calculation of the correlation volume in
Eq.~(\ref{eq:lnZ}), one must replace $\xi^d$ with $\xi^d\to\ell^d=\xi^d$. Consequently,
 $h(\tilde\xi/\xi)$, with $h(y)=y^{-\lambda}$ a general scaling function.
In essence, one is reducing the effective dimensionality such that $d\to
d^*=d-\lambda(a-1)$. $T-$linear resistivity obtains if $z=2-d^*$.  The
reduction in the effective dimensionality, $\lambda(a-1)$, can now be
fine-tuned so that $d^\ast\le 1$, thereby resulting in physically permissible
values of the dynamical exponent, $z\ge 1$.  Such fine scripting
of two length scales is also without basis at this time.

\section{Spectral Weight Transfer}

It is worth pointing out that non-commutative\cite{ncomm} field theories do permit $z<0$. 
Central to $z<0$ is UV-IR mixing and as a consequence a breakdown of the standard Wilsonian renormalization scheme.  While a field theory based on non-commuting coordinates is undoubtedly {\bf not} applicable to the cuprates, UV-IR mixing is certainly present as it has been documented experimentally both in the normal state\cite{cooper,uchida1,chenbatlogg,ncco,cuclo} as well as the superconducting state\cite{rubhaussen,marel,bontemps}.  Hence, the relevant question is: Can spectral weight transfer give rise to an additional length scale? 

\begin{figure}
\centering
\includegraphics[width=10.5cm]{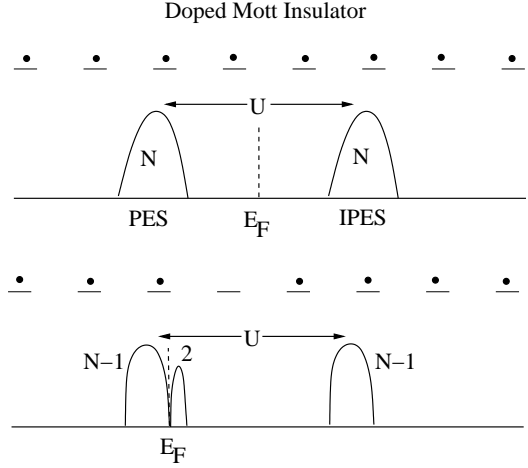}
\caption{Spectral weight transfer in a doped Mott insulator.  The Photoelectron
spectrum (PES) denotes the electron removal states while the electron-addition
states are located in the inverse photo-electron spectrum (IPES).  The on-site
charging energy is $U$. Removal of a single electron results in the creation
of two single particle states at the top of the lower Hubbard band. 
By state conservation, one state comes from the lower 
 and the other from the upper Hubbard band and hence spectral weight transfer across
the Mott gap. }
\label{fig2}
\end{figure}

To answer this question, we recount the well known argument\cite{sawa,eskes} on doping-dependent spectral-weight transfer across the Mott gap.  Consider 
the half-filled one-dimensional chain of one-electron atoms shown in
Fig. (\ref{fig2}) described by the Hubbard Hamiltonian,
\beq
H = -\sum_{i,j,\sigma} t_{ij}c_{i\sigma}^{\dagger}c_{j\sigma} + 
U\sum_{i} n_{i\uparrow}n_{i\downarrow}-\mu\sum_{i\sigma} n_{i\sigma},
\eeq
in which electrons hopping on a lattice between neighbouring sites with amplitude $t_{ij} = t\alpha_{ij}$ and chemical potential $\mu$ pay an energy cost $U$ anytime they doubly occupy the same site.  The operator $c_{i\sigma}$ ($c_{i\sigma}^\dagger$) annihilates (creates) an electron on site $i$ with spin $\sigma$ and $n_{i\sigma}$ is the occupancy on site $i$ with spin $\sigma$. 
When the hopping term vanishes, we can treat the half-filled system
as having one electron per site.   For a chain containing $N$ sites, there
are $N$ ways to remove an electron and $N$ ways to add an electron with an energy cost $U$. The operators that describe such excitations
are $\xi_{i\sigma}=c_{i\sigma}(1-n_{i-\sigma})$ which adds an electron to an unoccupied site and 
$\eta^\dagger_{i\sigma}=c^\dagger_{i\sigma}n_{i-\sigma}$ which creates a doubly occupied site.
Such operators create excitations in the lower and upper Hubbard bands, respectively, noted also in Fig.(\ref{fig2}) as the photo-electron and inverse photo-electron spectra.  When a single hole is created,  the number of ways to remove an electron is now $N-1$.  Surprisingly, the number of ways of adding an electron to the system so that the energy cost is $U$ also decreases to $N-1$.  This means that there are two less states at high and low energies.   Where are the remaining two states? They correspond to the two ways of occupying the empty site with spin up and spin down electrons.  Such states correspond to the addition part of the low-energy spectral weight and hence lie immediately above the chemical potential
as illustrated in Fig. (\ref{fig2}).   One of the states must come from the 
upper Hubbard band (UHB) as
 the high energy part
 now has a spectral weight of $N-1$, the other from the lower Hubbard band. Hence, for a single hole,
there is a net transfer of one state from high to low energy.  
 In general, 
simple state counting yields $2x$ for the growth of addition part of the low-energy spectral weight, $\Lambda(x)$ and $1-x$
 for the depletion of the high energy sector. In a Fermi liquid, adding holes simply creates quasiparticles near the chemical potential and hence cannot involve the high energy part.  In actuality, the dynamical contribution\cite{sawa,eskes}
to the LESW results in $\Lambda(x)>2x$.   The dynamical LESW corresponds to 
virtual excitations to the UHB.   Hence, in a strongly correlated system, some of the low-energy degrees of freedom arise from the high energy scale.  A bonus that can be explained from spectral weight transfer is the sign change of the Hall coefficient seen widely\cite{hall1,hall2,hall3} in the cuprates.  Particle-hole symmetry is restored in the lower Hubbard band when the addition and removal parts of the low-energy spectral weights are equal.  Equating the atomic ($U=\infty$) values for these weights, namely $1-x$ for the addition spectrum and $2x$ for the removal part, we obtain that the maximum doping level at which the Hall coefficient changes sign is 
$x=1/3$. This upper bound correlates with the experimentally observed
value for the vanishing of the Hall coefficient\cite{hall1,hall2,hall3} in the cuprates and is precisely the value obtained\cite{castillo,shastry} from strong-coupling calculations at large $U\approx\infty$.

\begin{figure}
\centering
\includegraphics[height=9.5cm]{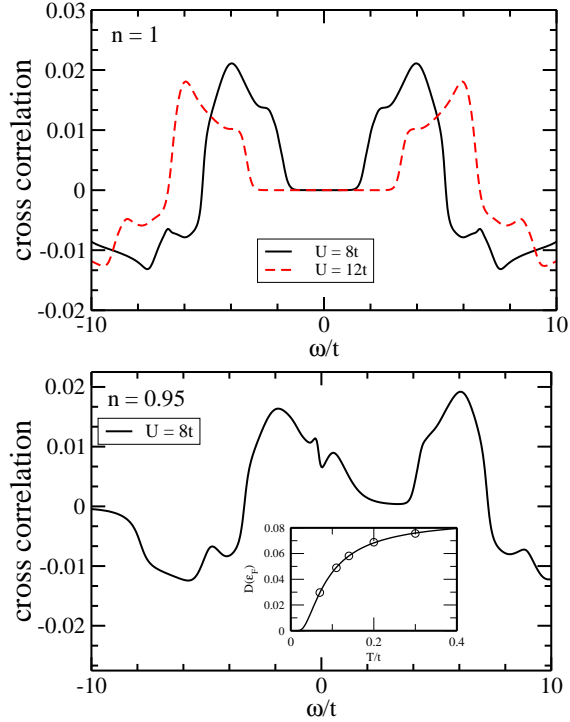}
\caption{Cross correlation or quantum interference $A_{\eta\xi}$ between the upper and lower Hubbard bands at half-filling, $n=1$ and at $n=.95$ at $T=0.1t$.
The dip at the chemical potential in the lower panel represents the pseudogap. The inset shows that this dip leads to a vanishing density of states at zero temperature and hence an orthogonality catastrophe.}
\label{crossterm}
\end{figure}

Spectral weight transfer described above can be reformulated in terms of the composite operators for the lower and upper Hubbard bands.   Write the electron operator as $c_{i\sigma}=\xi_{i\sigma}+\eta_{i\sigma}$ and the corresponding spectral function,  
\beq\label{green}
A(\vec k,\omega)&=& \frac{-1}{\pi}{\rm Im} FT(\theta(t-t')\langle \{c_{i\sigma}(t),c^\dagger_{j\sigma}(t')\}\rangle)\nonumber\\&=&A_{\eta\eta}+A_{\xi\xi}+2A_{\eta\xi},     
\eeq        
explicitly the imaginary part of the single particle Green function,
contains diagonal terms as well as a mixing term which carries the information regarding the interconnectedness between the high and low energy scales. It is from the mixing term that spectral weight transfer arises.  Shown in Fig. (\ref{crossterm}) is the mixing term computed using a self-consistent cluster method\cite{noncomm}). Regardless of filling, the cross term has both negative and positive contributions. This structure arises necessarily because 
the integral of 
$A_{\eta\xi}$ over all frequency 
\beq
\int_{-\infty}^{\infty} A_{\eta\xi}(k,\omega)d\omega=0
\eeq
yields the equal time correlator 
$\langle\{\xi_{\i\sigma},\eta^{\dagger}_{i\sigma}\}\rangle=0$, whose vanishing maintains the Pauli principle.  This implies that $A_{\eta\xi}$ is either zero, which it is not, or it must have both positive and negative parts, representing constructive and destructive interference, respectively, between different regions in energy.  At half-filling, upper panel, if particle-hole symmetry is present, the contributions
 of the cross term below and above the chemical potential
sum to zero independently, indicating that all energy scales need not be retained
to ensure the Pauli principle.   Such is not the case, however, 
at finite doping.  The lower panel in 
Fig. (\ref{crossterm}) indicates that a pseudogap develops at the chemical potential, indicating an orthogonality catastrophe.  Because a pseudogap subtracts spectral weight at low energies and transfers it to intermediate to high energy, the sum rule which ensures the Pauli principle is satisfied  only when $A_{\eta\xi}$ is integrated over all energy scales not simply up to the chemical potential (or some intermediate energy cutoff) as would be the case in projected 
models.  

That projection changes fundamentally the statistics of the particles arises from the truncation of the Hilbert space.  Consider the standard way of formulating the $t-J$ model from the Hubbard model: block diagonalise the Hubbard model into sectors with a fixed number of doubly occupied sites via a similarity transformation\cite{trans}, $S$, which connects sectors that differ by
a single double occupancy.  In such a scheme, the electron operator in the low-energy sector is transformed to
\beq\label{projec}
c_{i\sigma}\rightarrow Pe^{-S}c_{i\sigma}e^{S}P\equiv a_{i\sigma}
\eeq
where $P$ removes double occupancy.  To lowest order in $t/U$, $a_{i\sigma}=\xi_{i\sigma}$ and $\xi^\dagger_{i\uparrow}\xi^\dagger_{i\downarrow}=0$.  That is,
projected operators in  Eq. (\ref{projec}) block two states from being occupied rather one as would be the case for fermions.  Simply, projection of the
electron onto the lower Hubbard band does not result in a fermion. In fact, such projected objects are exclusons obeying generalized exclusion statistics\cite{haldane}. 

\begin{figure}
\centering
\includegraphics[width=7.5cm,angle=0.0]{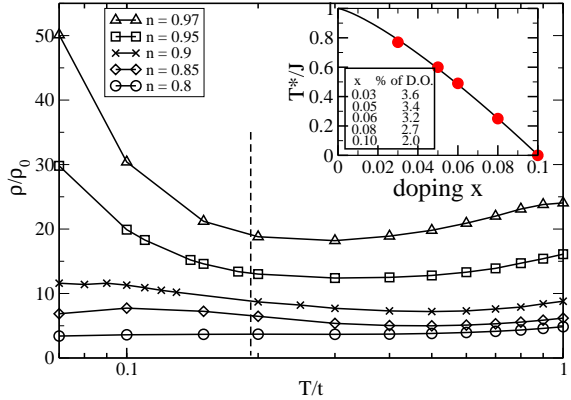}
\caption{Resistivity as a function of temperature for the Hubbard model (with $U=10t$) using the spectral function computed previously by Stanescu and Phillips\cite{fmott} for fillings $n=0.97, 0.95, 0.9, 0.85, 0.80$. $\rho_0=h/e^2$. The inset shows the doping dependence of the pseudogap energy scale, $T^\ast$.  The dependence obeys the functional form, $J(1-x\xi_{\rm do}^2/4a^2)$, where $\xi_{\rm do}$ is the average distance between doubly occupied sites, the percentage of 
which is indicated in the table.  }
\label{holeloc} 
\end{figure}

Since truncating the Hilbert space from four to three states per site changes the statistics, the limits $U\rightarrow\infty$ and $L\rightarrow\infty$ do not commute.  Whether or not this changes the physics fundamentally is open for
debate.   However, if transport is governed by a finite length scale for double occupancy, $\xi_{\rm do}$, then the correct order of limits is $L\rightarrow\infty$ followed by $U\rightarrow\infty$.  Projected models such as the $t-J$ model correspond
to the opposite regime, $U\rightarrow\infty$ and then $L\rightarrow\infty$.  In this case, $\xi_{\rm do}=\infty$.  A finite value of $\xi_{\rm do}$ would certainly function as the additional length scale that is required to describe T-linear resistivity.  In recent cluster calculations\cite{holeloc} on the Hubbard model, we observed an insulating state (see Fig. (\ref{holeloc})) below a critical doping level as is seen experimentally\cite{boeb1,boeb2,boeb3,boeb4,ando}. The pseudogap arises\cite{holeloc} from the eneryg gap between local cluster states with total spin $S$ and $S-1$. This energy difference is essentially the
triplet-singlet splitting and hence scales as $t^2/U$.
  We argued that the insulating state persists as long as $n_h\xi_{\rm do}^2<L^2$, $n_h=x(L/a)^2$ the number of holes. $n_h\xi_{\rm do}^2=L^2$ defines the percolation limit. By calculating the percentage of 
doubly occupied sites, we obtained $\xi_{\rm do}$ numerically and plotted the $T^\ast$-line, $J(1-cx(\xi_{\rm do}/a)^2)$, in the inset of Fig. (\ref{holeloc}). The agreement of this phenomenological fit with the resistivity data in which a metallic state obtains at $x=0.1$ lends credence to our assertion that $\xi_{\rm do}$ is the relevant length scale for the pseudogap.   Similar calculations on the $t-J$ model find a metallic state\cite{haule,prelovsek} consistent with the lack of commutativity of the limits $U\rightarrow \infty$ and $L\rightarrow\infty$.  But this question is far from settled.

\section{Unifying Principle: Zeros}

To help resolve the potential non-commutativity of $U\rightarrow\infty$ and $L\rightarrow\infty$, we seek a unifying principle from which spectral weight transfer, pseudogap phenomena, and broad spectral features emerge.  Such a principle does exist and stems from the analytical structure of the single-particle Green function whenever a gap is dynamically generated.  In Fermi liquids, the single-particle Green function has poles at the quasiparticle excitations. The divergence of the single-particle Green function defines the Fermi surface. In insulators, no such divergence obtains because quasiparticles are absent. However,
Luttinger's\cite{lutt} theorem
\beq\label{lutt}
\frac{N}{V}=2\int_{G(0,p)>0}\frac{d^3p}{(2\pi)^3}\equiv n_{\rm Lutt}
\eeq
has been argued\cite{dzy} to be still valid even when quasiparticles are absent.   The essence is that the integral defining the particle density is performed over a slice in momentum space where the Green function is positive.  While numerous approximate calculations\cite{essler} have found surfaces of zeros in 1D\cite{essler} or quasi-1D\cite{rice1,rice2} systems, it is unclear if the zeros are an artifact of the approximations.   Further, it is implicitly assumed
in the work of Dzyaloshinskii\cite{dzy} that the volume of the surface of
zeros equals the particle density. Should this not be the case, Eq. (\ref{lutt}) must be appended with a correction to recover the particle density.
 In fact, this asumption has been questioned for a specific model for a Kondo insulator\cite{rosch}. 

Recently, we have found a general proof\cite{proof} that allows us to identify precisely where the surface of zeros occurs regardless of spatial dimension provided certain symmetries obtain. We expand on that proof here.  To proceed, we note that the causal nature of the retarded Green function
\beq
G^{\rm ret}_\sigma(i,j,t,t')=-i\theta(t-t')\langle\{c_{i\sigma}(t),c_{j\sigma}^\dagger(t')\}\rangle
\eeq
permits it to be constructed entirely from the spectral function (that is, the imaginary part)
\beq
G^{\rm ret}_\sigma(\vec k,\omega)=\int_{-\infty}^\infty d\omega' \frac{ A_\sigma(\vec k,\epsilon')}{\epsilon-\epsilon'+i\eta},
\eeq
through the standard Hilbert representation.  For a Mott insulator, a gap of order $U$ occurs in the spectral function.  We will take the gap to have a width $2\Delta$ centered about $0$.  Within the gap, $A(\vec k,\epsilon)=0$.
This is a necessary condition for any gap.  Consequently, in the presence of a gap, the real part of the Green function evaluated at the Fermi energy 
reduces to 
\beq\label{real}
R_\sigma(\vec k,0)=-\int_{-\infty}^{-\Delta_-}d\epsilon'\frac{ A_\sigma(\vec k,\epsilon')}{\epsilon'}-\int_{\Delta_+}^\infty 
d\epsilon' \frac{A_\sigma(k,\epsilon')}{\epsilon'}.
\eeq
We now prove that when particle-hole symmetry is present, the retarded Green function
is an even function of frequency at the non-interacting Fermi surface.
As a consequence, Eq. (\ref{real}) is identically zero along that momentum surface. To proceed, we consider
 the moments
\beq
M^\sigma_n(k)\equiv\int \frac{d\omega}{2\pi} \omega^n d\omega G^{\rm ret}_\sigma(k,\omega)
\eeq
of the Green function.  For simplicity, we have set $\hbar=1$. Using the Heisenberg equations of motion,
we reduce\cite{moments} the moments in real space
\beq
M^\sigma_n(i,j)&=&\frac12\left[\langle\{[H,[H\cdots[H,c_{i\sigma}]\cdots]_{\rm n\, times},c_{j\sigma}^\dagger\}\rangle\right.\nonumber\\
&+&\left.\langle\{c_{i\sigma},[\cdots[c^\dagger_{j\sigma},H]\cdots H],H]_{\rm
  n\, times}\}\rangle\right]
\eeq
to a string of commutators of the electron creation or annihilation
operators with the Hubbard Hamiltonian.  The right-hand side of this expression is evaluated at equal times.  To evaluate the string of commutators, it suffices to focus on the properties of 
\beq
K_{i\sigma}^{(n)}=[\cdots[c_{i\sigma},H],\cdots H]_{\rm n \, times},
\eeq
where by construction, $K^{(0)}_{i\sigma}=c_{i\sigma}$.  We write the Hubbard Hamiltonian as $H=H_t+H_U$ where $H_U$ includes the interaction as well as the chemical potential terms. The form of the first commutator,
\beq
K_{i\sigma}^{(1)}=\sum_jt_{ij}c_{j\sigma}+Uc_{i\sigma}n_{i-\sigma}-\mu c_{i\sigma}
\eeq
suggests that we seek a solution of the form
\beq
K_{i\sigma}^{(n)}=\sum_j t_{ij}\Lambda_{j\sigma}^{(n)} +Q_{i\sigma}^{(n)}
\eeq
where $Q_{i\sigma}^{(n)}=[\cdots[c_{i\sigma},H_U],\cdots H_U]_{\rm n \, times}$ involves a string containing $H_U$ n times and in 
$\Lambda_{j\sigma}$, {\bf $H_t$ appears at least once}.  Our proof hinges on the form of $Q_{i\sigma}^{(n)}$ which we write in general as
$Q_{i\sigma}^{(n)}=\alpha_n c_{i\sigma}n_{i-\sigma}+\beta_n c_{i\sigma}$.
The solution for the coefficients
\beq
\begin{array}{ll} \alpha_{n+1}=(U-\mu)\alpha_n+U(-\mu)^n\\ \beta_n=(-\mu)^n
\end{array}
\eeq
is determined from the recursion relationship $Q_{i\sigma}^{(n+1)}=[Q_{i\sigma}^{(n)},H_U]$.  In the moments, the quantity which appears is 
\beq
\langle\{Q_{i\sigma}^{(n)},c^\dagger_{j\sigma}\}\rangle=\delta_{ij}\left[\alpha_n \langle n_{i-\sigma}\rangle+\beta_n\right]\equiv \delta_{ij}\gamma_n.
\eeq
Consequently, the moments simplify to
\beq\label{finalmoment}
 M^{\sigma}_n(i,j)=\delta_{ij}\gamma_n+\frac12\sum_lt_{il}\left(\langle\{
 \Lambda^{(n)}_{l\sigma},c^\dagger_{j\sigma}\}\rangle+h.c.\right).
\eeq

The criterion for the zeros of the Green function
now reduces to a condition on the parity of the right-hand side of 
Eq. (\ref{finalmoment}).  Consider the case of half-filling and nearest-neighbour hopping. Under these conditions, $\langle n_{i\sigma}\rangle=1/2$ and by particle-hole symmetry, $\mu=U/2$.  The expressions for $\alpha_n$ and $\beta_n$
lays plain that the resultant coefficients
\beq\label{gamma}
\gamma_n=\left(\frac{U}{2}\right)^n\frac{1+(-1)^n}{2}
\eeq
vanish for $n$ odd.  Consequently, $G_\sigma(k,\omega)$ is an even function if the second term in Eq. (\ref{finalmoment}) vanishes.  In Fourier space, the second term is proportional to the non-interacting band structure, which in the case of nearest-neighbour hopping is $t(k)=-2t\sum_{i=1}^d\cos k_i$.  As a result,
the Green function only has even moments at the momenta for which 
$\sum_{i=1}^d \cos k_i=0$. The vanishing of $t(k)$ defines the non-interacting 
Fermi surface.  Consequently,
the surface of zeros is pinned at the Fermi surface of the non-interacting system at half-filling {\bf whenever a Mott gap opens} in the presence of particle-hole symmetry.  This constitutes one of the few exact
results for Mott insulators that is independent of spatial dimension.  The only condition for the applicability of our proof is that the form of the gap
leads to finite integrals in Eq. (\ref{real}).  Hence, the minimal condition
is that the density of states vanishes at zero frequency as $\omega^\alpha$, $\alpha>0$.  Note, this encompasses even the Luttinger liquid case in which
$0<\alpha<1$.

To prove that this observation applies regardless of the range of the Coulomb interactions, we consider an argument directly from particle-hole symmetry.
Under a general particle-hole transformation, 
\beq\label{ph}
c_{i\sigma}\rightarrow e^{i\vec Q\cdot \vec r_i}c^\dagger_{i\sigma},
\eeq
with the chemical potential fixed at $\mu=U/2$ in the Hubbard model, the spectral function
becomes 
\beq\label{aspec}
A_\sigma(\vec k,\omega)=A_\sigma(-\vec k-\vec Q+2n\pi,-\omega),
\eeq
Hence, the spectral function is an even function of frequency for $\vec k=\vec Q/2+n\pi$. Consider one dimension and nearest-neighbour hopping.  In this case, the symmetry points are $\pm\pi/2$, the Fermi points for the half-filled non-interacting band.  In two dimensions, this proof is sufficient to establish the existence of only two points, not a surface of zeros.  To determine the surface, we take advantage of an added symmetry in higher dimensions.  For example, in two dimensions, we can interchange the canonical x and y axes leaving the Hamiltonian unchanged.  This invariance allows us to interchange $k_x$ and $k_y$ on the left-hand side of Eq. (\ref{aspec}) resulting in the conditions
\beq\label{crit1}
k_y=-k_x-q+2n\pi
\eeq
and by reflection symmetry
\beq\label{crit2}
-k_y=-k_x-q+2n\pi,
\eeq
where $\vec Q=(q,q)$.   For nearest neighbour hopping, the resultant
condition, $k_x\pm k_y=-\pi+2n\pi$, is the solution to $\cos k_x+\cos k_y=0$, which defines the Fermi surface for the non-interacting system.   When nearest-neighbour Coulomb interactions are present, the chemical potential changes to $(U+2V)/2$ at the particle-hole symmetric point and the proof follows as before.  Arbitrary density-density interactions and hence all Coulomb interactions are independent of $Q$ under a particle-hole transformation.  Hence, 
our proof applies regardless of the range of the Coulomb interactions.   A proposed proof\cite{tsvelik} for the existence of zeros in 1D relies explicitly on rotational invariance of the Green function.  Such an invariance is not applicable to the Hubbard model as the charge and spin velocities
are necessarily different as long as $U\ne 0$.  Hence, our proof constitutes the only proof for zeros exist in the Hubbard model regardless of spatial dimension.  Finally, existing calculations based on the random phase approximation in which zeros were found at the non-interacting Fermi surface in 1D\cite{essler} and quasi-1D\cite{rice1,rice2} systems are consistent with our proof.

While this proof applies strictly at half-filing with particle-hole symmetry, it places constraints on how the Fermi surface can evolve at finite doping in Mott systems. The consequences of this result are as follows.  First, in the presence of the Mott gap, the self-energy diverges
at zero frequency along the surface of zeros.  To prove this, observe that the single-particle
Green function can be written as $G_\sigma(k,\omega)=1/(\omega-\epsilon(k)-\Re\Sigma_\sigma(k,\omega)-i\Im\Sigma_\sigma(k,\omega))$,
where $\Sigma$ is the self-energy.  Near the Luttinger surface, $G\propto k-k_L$. $\Im\Sigma(k,0)\propto\delta(k-k_L)$. Note, however,
that $\Im G(k,\omega)=0$ for all energies within the gap.  Now simply invert the Green function to obtain that 
$\Re\Sigma_\sigma(k,\omega)\propto (k-k_L)^{-1}$, thereby proving the assertion.  Such a divergence is a clear indication that perturbation theory breaks down.  Zeros are a concrete manifestation of this breakdown.  The divergence of $\Re\Sigma$ prevents the renormalized energy band, $\epsilon(k)+\Re\Sigma_\sigma(k,0)$  from crossing the chemical potential, thereby resulting in an insulating state.  This reinterpretation of the Mott insulating state provides a general way of understand how insulating states arise through dynamical generation of a gap in the absence of 
symmetry breaking.  In dynamical cluster calculations,
the divergence of $\Sigma(\pi,0)$ \cite{jarrell} has been observed but attributed to antiferromagnetism.  What our proof makes clear is that the divergence of the self-energy is a result of Mottness itself; that is, it is not concomitant to $T=0$ ordering.

Second, the volume of the surface of zeros equals the particle density
only when particle-hole symmetry is present.  In proving this, we recall the
most general way of writing the electron density
\beq\label{charge}
\frac{N}{V}=n_{\rm  Lutt}+2\int\frac{d^dk d\omega}{(2\pi)^{d+1}}G(\vec k,\omega)\frac{\partial\Sigma(\vec k,\omega)}{\partial\omega}.
\eeq
Under Eq. (\ref{ph}), $G(\vec k,\omega)\rightarrow -G(\bf {-k-Q},-\omega)$ and $\Sigma(\vec k,\omega)\rightarrow -\Sigma(\vec k-\vec Q,-\omega)$, our Hubbard Hamiltonian remains invariant with the chemical potential fixed at the particle-hole symmetric value of $\mu=U/2$, and the additional
contribution to the electron density in Eq. (\ref{charge})
vanishes identically. For Fermi liquids, in which the self-energy is smooth,
the second term in Eq. (\ref{charge}) is integrated by parts\cite{luttinger}

\beq\label{charge}
 &\int d^d k\int\limits_0^\infty  d\omega \left( G(\vec k,\omega )\frac{\partial \Sigma(\vec k,\omega) }{\partial \omega} \right.&\nonumber\\ &\left. -G(-\vec k -\vec Q,\omega )\frac{\partial \Sigma(-\vec k-\vec Q,\omega )}{\partial \omega} \right)=0&
\eeq
 and shown
to be identically zero.   For a Mott system lacking particle-hole symmetry
but possessing a divergent self-energy, the standard integration by parts\cite{luttinger} of the second term in
Eq. (\ref{charge}) is not valid. The singular part of $G\partial_\omega\Sigma$
will always integrate to a non-zero value. That is, in the absence of particle-hole symmetry, no sum rule allows
us to equate the volume of the surface of zeros with the particle density as argued for perturbatively in a specific model for a Kondo insulator\cite{rosch}.

Third, Zeros are absent from projected models at half-filling.
Under projection, the real part of the
Green function reduces to the first integral in Eq. (\ref{real}) because projection does not preserve the contribution above the gap. This integral is always positive and hence zeros are absent.  Transforming the operators in the $t-J$ model to respect\cite{sashanew} the no double occupancy condition is of no help as the problem stems from
the loss of spectral weight above the gap once projection occurs. This result points to asymptotic slavery\cite{noncomm} of the Hubbard model or equivalently an example of the non-commutativity discussed earlier in that a low-energy reduction changes the physics.  Stated another way, the $t-J$ model violates Luttinger's theorem at half-filling.  The essence of this breakdown is that there is no kinetic energy in the $t-J$ model at half-filling, simply interacting spins. If there is no kinetic energy, there can be no surface of zeros. Such is not the case in the Hubbard model since no eigenstate of the Hubbard model has a fixed number of doubly occupied sites.

\begin{figure}
\centering
\includegraphics[width=8.5cm]{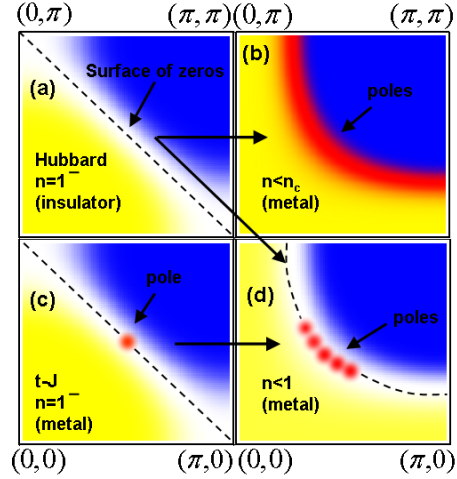}
\caption{Evolution of the surface of zeros in the first quadrant of the FBZ. Yellow indicates $\Re G>0$ while blue $\Re G<0$. The Hubbard model is constrained to have a surface of zeros as $n\rightarrow 1$ whereas the t-J model is not. The two options upon doping represent
weak-coupling (1a or c to d) and strong-coupling (1a to b). The transition from (a) to (b) requires a critical point at $n_c$ whereas (d) does not. Experiments\cite{boeb1,ando} indicating an insulating
state for $n>n_c$ are consistent with an abrupt transition from (a) to (b).}
\label{options}
\end{figure}

Fourth, even at infinitesimal doping, the t-J and Hubbard models are not equivalent.  Strictly speaking, at half-filling, it is not entirely appropriate to compare the $t-J$ and Hubbard models because the former has no spectral weight above the chemical potential.  Perhaps the proper way is to compare 
the two models in the limit of infinitesimal doping (see Figs. (\ref{options}a) and (c)). For one hole ($n=1^-$), numerical and analytical studies
\cite{z1,z2,z3} find
 a quasiparticle in the t-J model with weight $J/t$ at $(\pi/2,\pi/2)$ whereas in the Hubbard model\cite{sorella}, the quasiparticle weight vanishes as $Z\propto L^{-\theta}$, $\theta>0$, $L$ the system size. This result is consistent because the one-hole system should closely mirror the half-filled system, which for the Hubbard model must have
a surface of zeros for $n=1$ while no such constraint exists in the t-J model. This has profound consequences for the minimal model needed to describe the cuprates. 

Fifth, band structure cannot affect the existence of zeros
in the presence of the Mott gap.
At present,
our proof applies to any kind of band structure that is generated from
hopping processes which remain unchanged  after the application of Eq. (\ref{ph}).  In general, the two kinds of hopping processes 
transform as $\epsilon(\pi-k_x,\pi-k_y)=-\epsilon(k_x,k_y)$ and $t'(\pi-k_x,\pi-k_y)=t'(k_x,k_y)$.  The latter is relevant to the cuprates.
If only such hopping is present, the
surface of zeros is no longer the diagonal $(\pi,0)$ to $(0,\pi)$ (or the point
$\pm\pi/2$ in 1D),
but rather the ``cross'' $(0,\pi/2)$ to $(\pi,\pi/2)$ and $(\pi/2, 0)$ to $(\pi/2,\pi)$ (or, in 1D, the
points $\pm\pi/4$ and $\pm 3\pi/4$).  When both types of hopping are present, no symmetry arguments can be made. However, at this point it is imperative that we distinguish between the 1D and 2D cases.  In 1D, if only two Fermi points exist, the Luttinger theorem requires that the zeros be at the symmetric Fermi points, $\pm\pi/2$.  But in principle there can be 2n Fermi points in 1D. Consider the case of $t$ and $t'$ hopping.  Four Fermi points exist for $t'>t/2$.  In such cases, the zeros need not be at the Fermi surface of the non-interacting system.  Such 1D cases share the complexity of the general case of an asymmetric band structure in 2D.  When this state of affairs obtains, we can establish the existence of zeros by a key assumption:  the Green
function is a continuous function of the hopping parameters $t$ and $t'$.  When only
$t$ is present, $R_\sigma(k,0)$ has one sign (plus) near $k=(0,0)$
(or, in 1D, k=0), and the opposite (minus) near 
$k=(\pi,\pi)$ ($k=\pi$ in 1D)
and will vanish on the zero line. Alternatively, if we have $t'$ hopping,
$R_\sigma(k,0)$
will have a certain sign near $k=(0,0)$ and $k=(\pi,\pi)$, 
and the opposite sign
near $k=(0,\pi)$ and $k=(\pi,0)$ and will vanish on the "cross". 
From continuity, for $t'\ll t$, $R_\sigma(k; t, t')$ will have the same 
sign structure as $R_\sigma(k;
t, t'=0)$. That is, it will change sign when going from $(0,0)$ to $(\pi, \pi)$
regardless of the path taken.  Therefore, the line of zeros exists
for small enough t', the relevant limit for the cuprates. In the opposite limit, $t'\gg t$, a similar argument holds. 

Finally, the surface of zeros defines the pseudogap at finite non-zero doping.
  At finite hole-doping ($x=1-n$), the chemical potential jumps into the lower Hubbard band.  For $x\approx 1$, $R_\sigma(k,0)$ must vanish on some momentum 
surface because  most of the spectral weight at $(\pi,\pi)$ still resides above the chemical potential, whereas at $(0,0)$ it resides below.    Consequently, a surface of zeros should still be present so long as a gap, commonly known as the pseudogap, opens up to some doping $x_c$.  Satisfying the zero condition, Eq. (\ref{real}), requires spectral weight to lie immediately above the chemical potential.  Spectral weight
 transfer across the Mott gap depicted in Fig. (\ref{fig2}) and discussed extensively previously mediates the zeros.  
  The pseudogap that results is necessarily asymmetrical for
 small $x$ because the $2x$ peak must lie closer to the chemical potential than the $1-x$ contribution in order for Eq. (\ref{real}) to vanish.   Hence, an asymmetric pseudogap is a direct indication that Mottness rather than weak-coupling/ordering physics is the operative cause.  In fact, such an asymmetry has been observed in the pseudogap regime in the cuprates\cite{asym}.  The real question is how do the zeros evolve upon doping.
 As shown in Fig. (\ref{options}), either the surface of zeros vanishes
abruptly above a critical doping or it shrinks smoothly.  In either case, the pseudogap must vanish at a finite number of $k-$points.  A hard gap would involve all momenta.  Hence, the pseudogap must vanish as $\omega^\alpha$, with $\alpha>0$ . Because poles cannot exist within an $\epsilon$ neighbourhood of a zero if the pseudogap arises from Mottness, quasiparticles cannot survive at the nodes.   Weak-coupling or Fermi liquid-type scenarios permit nodal quasiparticles. Numerics\cite{stanescu3,c2,c3} point to such a non-symmetry breaking pseudogap in the Hubbard model at finite doping whose cause is Mottness.  Also, Stanescu and Kotliar\cite{stanescu4} have shown convincingly using a cellular dynamical cluster method on the Hubbard model with nearest-neighbour hopping that
the spectral features of a momentum-dependent pseudogap are directly
related to the surface of zeros.  This is the first extensive numerical study
that has established the hard connection between zeros and the pseudogap. 

\section{Concluding Remarks}

Because of its explanatory power, the surface of zeros can be argued to play a role in doped Mott insulators analogous to the Fermi surface in Fermi liquids.  However, the only
feature they share in common, is their volumes when the Hamiltonian has particle-hole symmetry.   In Fermi liquids, no spectral weight transfer occurs, no pseudogap phenomena exists and quasiparticles are well defined.  Just the opposite obtains in doped Mott systems. 
That broad spectral features naturally follows from the surface of zeros is clear because sharp quasiparticles require the renormalized energy band to cross the chemical potential. This is not possible as long as $\Re\Sigma_\sigma(k_L,0)$ diverges as has been demonstrated.   Hence, one of the most nettling problems with the cuprates, namely broad spectral features\cite{norman} has a natural resolution on the surface of zeros. The zeros also offer a concrete realisation of the non-commutativity of $U\rightarrow\infty$ and $L\rightarrow\infty$ as they are present in the Hubbard model at half-filling but not in the projected $t-$J model. A vanishing of the surface of zeros above a critical doping level
 will define a length scale associated with the gap.  Such a length scale will undoubtedly play a role in the transport properties in the strange metallic regime.  A complete treatment of the transport properties and the nature of the excitations on the surface of zeros will necessitate a complete field theory of Mottness. 

\section{Acknowledgements} I wish to acknowledge several of my students, T. Stanescu, T.-P. Choy, and D. Galanakis whose work led to many of the ideas presented here.
I also acknowledge the collaboration with Claudio Chamon which
 led to the proof on $T-$linear resistivity.  I also thank 
E. Fradkin whose insistence that zeros did not exist forced us to construct the proof on zeros in Sec. 4 and A. Tsvelik for an e-mail exchange
 on the 1D zero problem.  This work was funded by the
 NSF Grant No. DMR-0605769.  The Workshop on Mottness and Quantum
 Criticality was made possible by a grant from NSF and the DOE.


\begin{thebibliography}{99}
\bibitem{enscal}In systems with a Fermi surface, it is the Fermi energy that replaces the plasma frequency in Eq. (\ref{drude}). Transport in such systems is also incompatible with one-parameter scaling.
\end{thebibliography}
\end{document}